\documentclass[iop]{emulateapj}
\usepackage{graphicx,color}
\usepackage{comment}
\usepackage{hyperref}
\usepackage{url}
\usepackage{breakurl}
\shortauthors{Allers \etal}
\shorttitle{PSO J318.5-22 is a $\beta$ Pic member}

\begin{document}

\newcommand{\Ks}{\mbox{$K_S$}}
\newcommand{\Lp}{\mbox{$L^{\prime}$}}
\newcommand{\mbol}{\mbox{$m_{\rm{bol}}$}}
\newcommand{\degs}{\mbox{$^{\circ}$}}
\newcommand{\perpix}{\mbox{pixel$^{-1}$}}
\newcommand{\mjup}{\mbox{$M_{\rm{Jup}}$}}
\newcommand{\rjup}{\mbox{$R_{\rm{Jup}}$}}
\newcommand{\etal}{et al.}
\newcommand{\eg}{e.g.}
\newcommand{\ie}{i.e.}
\newcommand{\htwoo}{{\hbox{H$_2$O}}}   
\newcommand{\kms}{\mbox{km~s$^{-1}$}} 
\newcommand{\vsini}{\mbox{$v$~sin($i$)}} 
\newcommand{\teff}{\mbox{$T_{eff}$}} 
\newcommand{\logg}{\mbox{log($g$)}} 

\newcommand{\pso}{PSO~J318.5$-$22}
\newcommand{\bpic}{$\beta$~Pic}

\newcommand\note[1]{{#1}}

\def\substitute@option#1#2{%
 \ClassWarning{aastex}{%
  Substyle #1 is deprecated in aastex.
  Using #2 instead (please fix your document).
 }\@nameuse{ds@#2}%
}%
\newcommand\ionscript[2]{#1$\;${\tiny\rmfamily{#2}}\relax}%
\newcommand\ionfootnote[2]{#1$\;${\scriptsize\rmfamily{#2}}\relax}%

\title{\note{The Radial and Rotational Velocities of PSO J318.5338$-$22.8603}, a Newly Confirmed Planetary-Mass Member of the $\beta$~Pictoris Moving Group}

\author{K.~N.~Allers}
\affil{Department of Physics and Astronomy, Bucknell University, Lewisburg, PA 17837, USA; k.allers@bucknell.edu}

\author{J.~F.~Gallimore}
\affil{Department of Physics and Astronomy, Bucknell University, Lewisburg, PA 17837, USA}

\author{Michael~C.~Liu} 
\affil{Institute for Astronomy, University of Hawaii, 
2680 Woodlawn Drive, Honolulu, HI 96822, USA}

\author{Trent~J.~Dupuy} 
\affil{The University of Texas at Austin, Department of Astronomy, 2515 Speedway C1400, Austin, TX 78712, USA}

\begin{abstract}

PSO J318.5338$-$22.8603 is an extremely-red planetary-mass object that has been identified as a candidate member of the $\beta$ Pictoris moving group based on its spatial position and tangential velocity.  We present a high resolution $K$-band spectrum of PSO J318.5338$-$22.8603.  Using a forward-modeling Markov Chain Monte Carlo approach, we report the first measurement of the radial velocity and \vsini\ of \pso, $-$6.0$^{+0.8}_{-1.1}$~\kms\ and 17.5$^{+2.3}_{-2.8}$~\kms, respectively.  We calculate the space velocity and position of \pso\ and confirm that it is a member of the $\beta$~Pictoris moving group. Adopting an age of 23$\pm$3~Myr for \pso, we determine a mass of $8.3\pm0.5$~\mjup\ and effective temperature of $1127^{+24}_{-26}$~K using evolutionary models. \note{PSO J318.5338$-$22.8603 is intermediate in mass and temperature to the directly-imaged planets $\beta$~Pictoris~b and 51~Eridani~b, making it an important benchmark object in the sequence of planetary-mass members of the $\beta$~Pictoris moving group.}
Combining our \vsini\ measurement with recent photometric variability data, we constrain the inclination of \pso\ to $>29^{\circ}$ and its rotational period to 5--10.2~hours.  The equatorial velocity of \pso\ indicates that its rotation is consistent with an extrapolation of the velocity-mass relationship for solar system planets.

\end{abstract}


\section{Introduction}

Objects with masses below the hydrogen burning limit \citep[$\sim$75~\mjup;][]{burrows01} cool as they age, resulting in a mass-age-luminosity degeneracy.  Several-Gyr-old brown dwarfs can have the same luminosities as much younger, several-Myr-old planetary-mass ($\lesssim$13~\mjup) objects.  Thus, to determine the mass of a single substellar object, one must have an independent \note{constraint on its} age.

As planetary-mass objects and brown dwarfs cool, they also contract, increasing their surface gravity as they evolve.  Thus evidence of low gravity (\eg\ weak FeH and alkali features, strong VO absorption) in a spectrum can \note{corroborate} the youth of an object.  As noted in \citet{allers13}, however, spectroscopic ages are not particularly accurate, and objects of the same age and spectral type can have different indicators of youth.

\note{Another method to} determine the age of a substellar object is via association with a young stellar moving group. 
These groups have ages known from the stellar members of the groups, either from model isochrone fitting of the members on an HR diagram \citep[\eg][]{zuckerman01,pecaut12} or the location of the lithium depletion boundaries \citep[\eg][]{king00,binks14}.  Though the estimated age of an individual group may vary by up to a factor of $\sim$2 depending on the method of age determination, the various methods agree on the relative ages of the groups \citep[\eg][]{mentuch08,dasilva09,bell15}.  In order to link an object to a local young moving group, its galactic kinematics ($UVW$) and position ($XYZ$) must agree with the other known members of the group.

PSO J318.5338$-$22.8603 (hereinafter \pso), discovered by \citet{liu13}, is the first free-floating object with colors and luminosity that overlap the young, dusty, \note{planetary-mass companions} to HR~8799 and 2MASS~J1207-39.  Based on its parallax, proper motion, and spectroscopic signatures of youth (low-gravity), \citeauthor{liu13} identified \pso\ as a candidate member of the $\beta$~Pictoris (hereinafter \bpic) moving group.  \citeauthor{liu13} estimate a mass of 6.5$^{+1.3}_{-1.0}$~\mjup\ for \pso, based on evolutionary models for an isochronal age of 12$^{+8}_{-4}$~Myr and the luminosity of \pso.  Here, we report a determination of the radial and rotational velocities of \pso\ for the first time and assess the probability that \pso\ is a member of the \bpic\ moving group.  We re-estimate the physical properties of \pso\ based on recent determinations of the age of the \bpic\ moving group \citep[23$\pm$3~Myr;][]{mamajek14}.  \note{With a mass and effective temperature intermediate to the directly-imaged exoplanets $\beta$~Pictoris~b \citep{lagrange09} and 51~Eridani~b \citep{macintosh15}, \pso\ is an important free-floating analog to these young planets.}  In addition, combining our measurement of \vsini\ with the recent measurements of \pso's orbital period \citep{biller15}, we constrain its spin axis inclination and equatorial velocity.

\section{Observations and Data Reduction}

We obtained a high resolution (R~$\approx$~14000) spectrum of \pso\
using the GNIRS spectrograph \citep{elias06} on the Gemini-North
Telescope.  We used the 111~lines~mm$^{-1}$ grating with a 0.15''
slit, resulting in a 2.27--2.33~$\mu$m spectrum.  Our queue
observations were obtained on UT 26 November 2013 under clear skies
while the seeing was 0.49''.  Seven exposures of 600~s were obtained,
nodding 6\arcsec\ along the slit between exposures for a total integration
time of 70 minutes.  Immediately after our observations of \pso, we
obtained flat field and Xenon-Argon lamp images.
We also observed an A0 star, HIP~110746, as a telluric standard using the same instrument
setup.  

Data were reduced using the {\tt REDSPEC} reduction
package to spatially and spectrally rectify each exposure, referencing the Keck/NIRSPEC Echelle Arc Lamp Tool (\url{http://www2.keck.hawaii.edu/inst/nirspec/N7arclamps.html}) to identify the wavelengths of lines in our arc lamp spectrum.
After nod-subtracting pairs of exposures, we create a spatial profile which is the median intensity across all wavelengths at each position along the slit.  
To remove any residual sky emission lines from our nod-subtracted pairs and to determine the noise per pixel at each wavelength, we identify pixels in the spatial profile that do not contain significant source flux.  At each wavelength, we fit a line to the intensity of these background pixels vs.~position along the slit.  
The fitted line provides the background level as a function of position and the scatter of the background pixels about the line gives the noise per pixel.\footnote{This assumes that the shot noise is dominated by the background.  \pso\ is faint, so the sky background level is $>$3 times brighter than the flux of our source across all wavelengths of our spectrum.}
We then performed a
profile-weighted extraction \citep{horne86} on each exposure and combined the resulting spectra
 using a robust weighted mean with the {\tt xcombspec} procedure from the SpeXtool package
\citep{cushing04}.  The median signal-to-noise of our final spectrum is 21.

We determine the FWHM of the line spread function (LSF) by fitting 1-dimensional gaussians to the Xenon and Argon lines in an extracted spectrum of the arc lamp image.  For the seven arc lamp lines, we calculate a mean and standard deviation for the LSF~FWHM of of 0.000163$\pm$0.000015~$\mu$m, corresponding to a resolution, $R = \lambda/ \Delta \lambda = 14100$.  

\section{Determining Radial and Rotational Velocities}

Unfortunately, no radial velocity standard was observed as a part of
our Gemini program.  We
use an approach similar to that described by \citet{blake10},
wherein we employ forward modeling to simultaneously fit the wavelength solution of our spectrum, the rotational and radial velocities of \pso, the scaling of telluric line depths, and the FWHM of the instrumental LSF.

\subsection{Creating a Forward Model}
Our observed spectrum of \pso\ is the combination of the spectrum of the
brown dwarf, telluric absorption, and effects
from the instrument (wavelength-dependent response and line spread functions).  We
create myriad model spectra and evaluate the best fit parameters by comparison to our
observed spectrum of \pso.  The components of our forward model are:

\begin{itemize}
\item{{\bf Wavelength Solution:}
When reducing our data, the wavelength solution of our observed spectrum is determined from seven Xenon-Argon arc lines with wavelengths of 2.29-2.33~$\mu$m.  There are no arc lines for the 2.27--2.29 $\mu$m portion of our spectrum.  In addition, the typical residuals to our wavelength solution for each arc line correspond to $\sim$5~km~s$^{-1}$.  
Our spectrum itself contains many telluric absorption features which allow us to refine the wavelength solution as a part of our forward modeling.
We calculate the wavelength solution of our observed spectrum using a
2nd-degree polynomial.  Rather than having the coefficients of the
polynomial as our fitted parameters, we use barycentric Lagrangian interpolation \citep{berrut04, waring79}, solving for the wavelengths of the first,
middle, and last pixels of the spectrum and determining the
coefficients of the unique 2nd-degree polynomial that will fit these three
points.  We start by adopting the wavelengths of our reduced spectrum as the initial guess.}  

\item{{\bf Brown Dwarf Atmosphere Model:}
We use the BT-Settl model atmospheres \citep{allard12} as the intrinsic
spectrum of \pso\ in our forward modeling.  The BT-Settl models are provided for two different
assumed solar abundances (Asplund et al. 2009 [AGSS] and Caffau et
al. 2011 [CIFIST]).  We use model grids with effective
temperatures (\teff) of 1100--2500~K and surface gravities (\logg) of
3.5--5.5~dex (where $g$ is in units of cm~s$^{-2}$.)   We adjust the wavelengths of the model atmosphere
to account for the radial velocity of the object, $\lambda =
\lambda_0 +( (v_r + v_{bary}) \lambda_0/c)$, where $v_r$ is the radial velocity of \pso, and $v_{bary}$ (28.1 \kms) is the barycentric motion projected away from \pso\ at the time of our observation.   
We then convolve the model spectrum to account for rotation, \vsini, using a
convolution kernel created by the {\tt LSF\_ROTATE} task\footnote{A part
of the IDL Astronomy User's Library}.  Our velocity-shifted, rotationally broadened model atmosphere approximates the spectrum of the brown dwarf itself.}

\item{{\bf Telluric Absorption:}
We use the telluric spectrum of
\citet{livingston91}\footnote{available for download at
  \url{ftp://ftp.noao.edu/catalogs/atmospheric_transmission/transdata_1_5_mic.gz}}.
We adjust the
depths of the telluric features and create a new telluric spectrum, $T(\lambda) =
T_0(\lambda)^\tau$, where $T_0(\lambda)$ is the
\citet{livingston91} spectrum, and $\tau$ is one of the parameters of
our fit.   The velocity-shifted, rotationally broadened brown dwarf model atmosphere is then multiplied by the telluric spectrum, approximating the spectrum of \pso\ as it enters the telescope.}

\item{{\bf Instrumental Effects:}
We approximate the instrumental LSF as a gaussian and fit for the FWHM of the
gaussian convolution kernel.  We convolve the brown dwarf model atmosphere with telluric absorption by this gaussian kernel.  
We account for any changes in spectral shape between our model and the observed spectrum by taking the ratio of our observed spectrum of \pso\ to
our model spectrum, fitting the smoothed ratio with a 3rd order
polynomial, multiplying our model by the fit, and then scaling our model to
the median of our observed spectrum.}
\end{itemize}

In total, our forward model has nine free parameters:  the \teff\ and \logg\ of the atmospheric model, the $v_r$ and \vsini\ of \pso, $\tau$ for the telluric spectrum, the LSF FWHM, and the wavelengths of the first, middle and last pixels.  The forward model is compared to our observed spectrum, and the parameters used to create the forward model are adjusted to achieve the best fit.

\subsection{Parameter Optimization with AMOEBA}

{\tt AMOEBA} uses a downhill simplex method \citep{nelder65} to find the minimum of a multivariate function.  We used IDL's implementation of {\tt AMOEBA}, which is based on \emph{Numerical Recipes in C} \citep{numericalrecipes}.   We run our {\tt AMOEBA} optimization for each model atmosphere having \teff\ = 900 - 2500~K and \logg\ = 3.5-5.5 in the CIFIST and AGSS grids.  {\tt AMOEBA} does not conduct a thorough search of parameter space, so it is important to seed it with reasonable parameter estimates.  We initialize {\tt AMOEBA} with $v_r$ = 0~\kms\ and \vsini\ = 15~\kms, typical of observed values of ultracool dwarfs \citep{blake10}.  We use the FWHM of the LSF as determined from our arc lamp spectrum (\S 2) as a starting point.  To get a starting point for $\tau$, we divide our observed telluric standard spectrum by an LSF-convolved telluric absorption spectrum and find the value of $\tau$ (0.7) that, by eye, results in a featureless spectrum (as expected for an A0 star from 2.27--2.33~$\mu$m).

\begin{figure}
\centerline{\includegraphics[angle=270, width=3.5in]{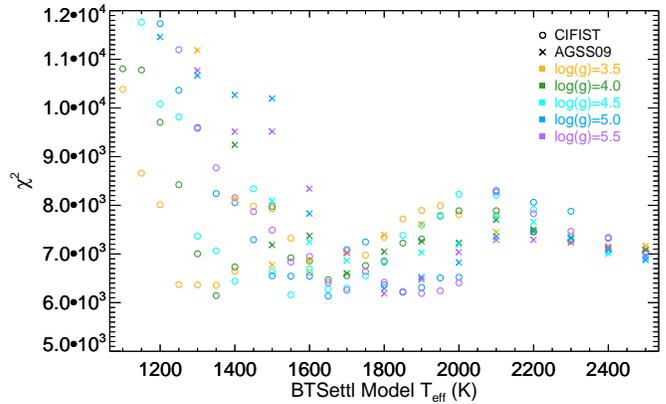}}
\caption[$\chi^2$ for Various Atmospheric Models]
{\label{fig:amoeba_chi} {\tt AMOEBA}'s best fit $\chi^2$ using BT-Settl models.  The minimum $\chi^2$ occurs for the CIFIST model with $T_{eff}$=1650~K and log($g$)=5.0.  In general, the CIFIST models provide marginally better fits than the AGSS09 models.}
\end{figure}

We find that the atmospheric model giving the best overall fit (minimum $\chi^2$), is the CIFIST model having \teff\ = \note{1650}~K and \logg\ = \note{5.0}, in reasonable agreement with the temperature (1400--1600~K) and log(g) (4.0--4.5) of the atmospheric models which fit the low-resolution 0.8--2.5~$\mu$m spectrum of \pso\ \citep{liu13}.  The best fit parameters returned for this model are: \note{FWHM of LSF=0.000169~$\mu$m, $\tau$=0.868, \vsini=11.6~\kms, and $v_r$=-7.4~\kms.}

Overall, the CIFIST models provide better fits than the AGSS models (Figure \ref{fig:amoeba_chi}).  
Regardless of the assumed solar abundance, \teff, or \logg\ of the atmospheric model, we find that {\tt AMOEBA} converges on a radial velocity  \note{of -7.4 to -2.3 \kms\ }.  In contrast, the \vsini\ that {\tt AMOEBA} finds varies significantly (0 to 80 \kms) depending on the model used.  
We find that the fitted \vsini\ and resulting $\chi^2$ of the model fit are correlated.  For atmospheric models with fits having higher $\chi^2$ values, {\tt AMOEBA} converges on higher values of \vsini.  From this analysis, it is unknown if the higher $\chi^2$ values mean the \vsini\ of \pso\ is genuinely low, or if for atmospheric models that do not approximate the spectrum of \pso\ a higher \vsini\ smears out the lines and results in a better fit.

Our {\tt AMOEBA} fitting provides a useful first estimate of the best fit parameters but \note{lacks a straightforward way to determine their uncertainties}.  In addition, the correlation between $\chi^2$ and \vsini\ motivates a more sophisticated analysis to determine our best fit parameters and their associated uncertainties.

\subsection{Parameter Optimization with MCMC and DREAM(ZS)}
\begin{figure}
\centerline{\includegraphics[width=3.8in]{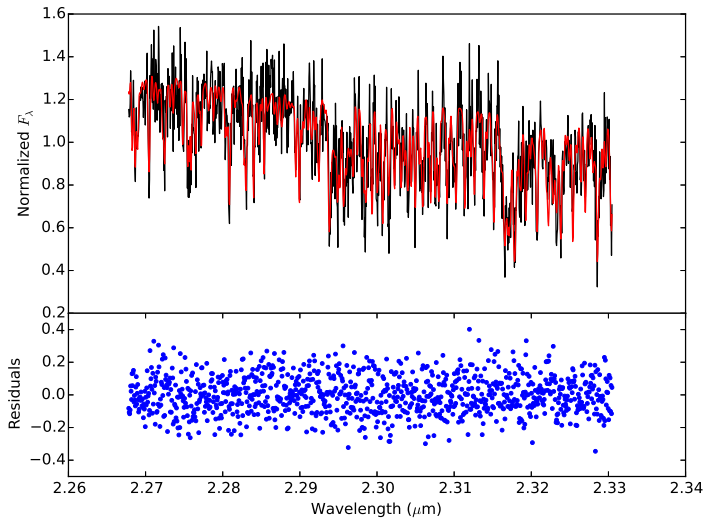}}
\caption[Best Fit Model]
{\label{fig:model_fit} Our observed spectrum of \pso\ (black) compared to our the forward model with our best fit parameters (red).  The median residual of the fit is 0.079 (in normalized F$_{\lambda}$ units), which is larger than the median uncertainty (0.048) of our observed spectrum.  The rms of the residuals (0.12) indicates systematic uncertainties of $\approx$10\%.}
\end{figure}

\begin{figure*}
\centerline{\includegraphics[width=6in]{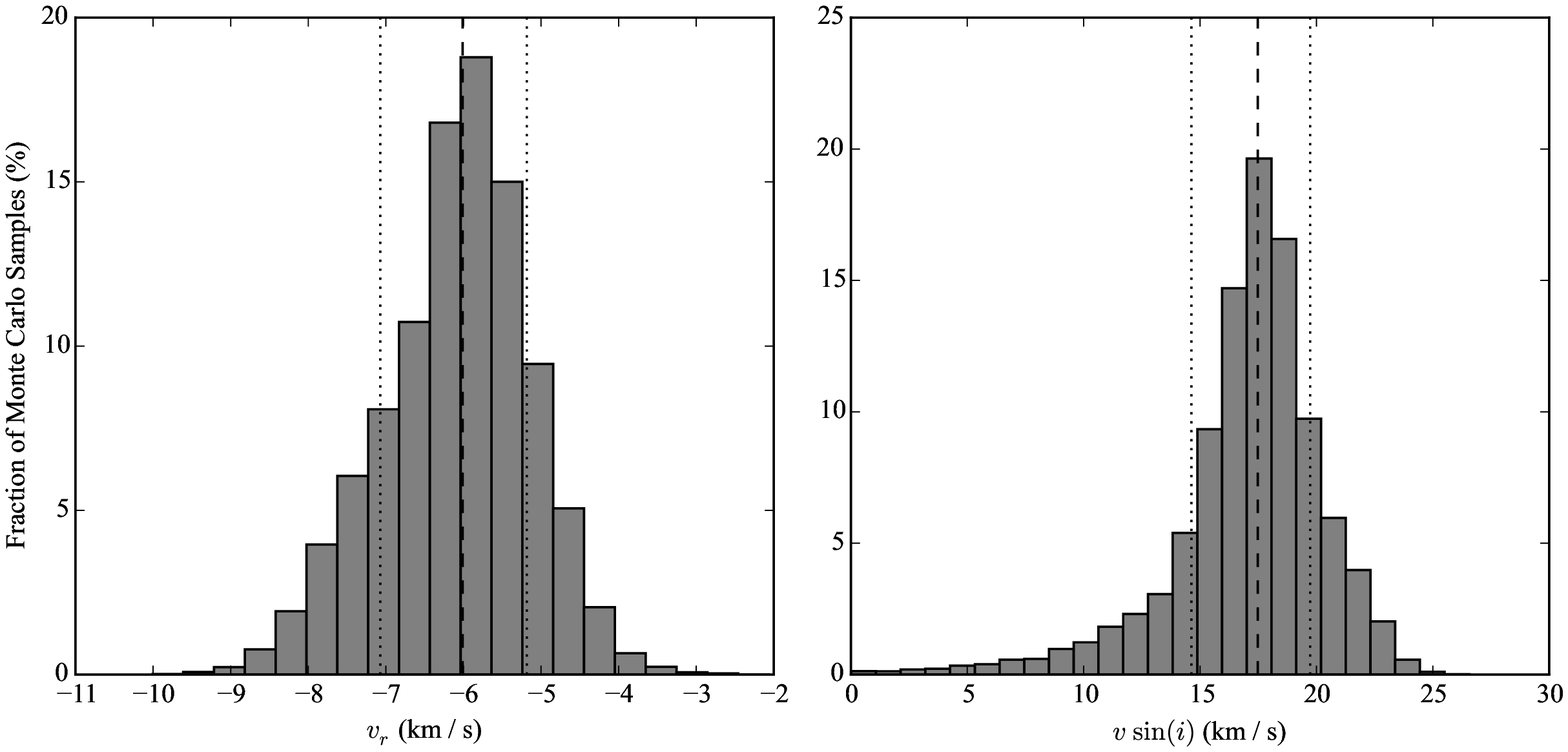}}
\caption[MCMC Posteriors]{\label{fig:dreamz_fits}Histograms of the \note{marginalized} posterior distributions for $v_r$ and \vsini\ from our MCMC analysis.  The dashed lines indicate the median of each distribution.  The dotted lines indicate the 68\% confidence limits (1-$\sigma$) centered on the median.}
\end{figure*}

To better determine the best fit parameters of our forward model as well as their 
\note{marginalized} posterior distributions, we use a Markov Chain Monte Carlo (MCMC) approach.  We create forward models in the manner described above but in addition allow for a continuous distribution of \teff\ and \logg\ by linearly interpolating between atmosphere models in the CIFIST grid. We
employ the DREAM(ZS) algorithm of \citet{terbraak08}, which
involves an adaptive stepper, updating model parameters based on chain
histories. A brief description of the algorithm follows. 

Three chains are run in parallel, and each chain is a history of
accepted parameter vectors $\mathbf{p}$. 
Current values of the 3 chain vectors are stored in the array $\mathbf{X}$, and chain histories are stored in the array $\mathbf{Z}$. 
To simplify the notation,
$\mathbf{X}[i]$ refers to the $n$-dimensional, $i$-th parameter vector
currently stored in $\mathbf{X}$ (i.e., the $i$-th chain in
$\mathbf{X}$), and we adopt the same notation for the history array,
$\mathbf{Z}[i]$.  We initially seed each chain with the parameters of the best-fit model as determined by our {\tt AMOEBA} analysis (\S 3.2).

At each iteration and for each chain $\mathbf{X}[i]$, two parameter
vectors $\mathbf{Z}[j]$ and $\mathbf{Z}[k]$ are selected at random and
without replacement from $\mathbf{Z}$. A trial parameter vector,
$\mathbf{x}$, is then calculated by the difference,
\begin{equation}
\mathbf{x}[i] = \mathbf{X}[i] + (1 + \mathbf{e}) \gamma
(\mathbf{Z}[j] - \mathbf{Z}[k]) + \mathbf{\epsilon},
\end{equation}
where $\gamma$ is a constant of order unity, $\mathbf{e}$ is an
$n$-dimensional vector of uniformly-distributed random numbers spanning
$[-0.05, 0.05]$, and $\mathbf{\epsilon}$ is a vector of
normally-distributed random numbers with zero mean and $\sigma =
10^{-6}$. 

The trial vector is accepted or rejected using the standard
Metropolis-Hastings likelihood ratio criterion \citep{metropolis53, hastings70}. We assume normally distributed
uncertainties. Therefore, the log-likelihood of a parameter vector is
given by $\ln(\mathbf{p}) = C -\chi^2 / 2$, where $C$ is a normalization
constant that cancels out in the likelihood ratio, and $\chi^2$ is the
usual chi-squared statistic.

Every ten iterations (the thinning interval), the current chain values
$\mathbf{X}$ are appended to the history array $\mathbf{Z}$. Note that
once the chains are appended to $\mathbf{Z}$, they lose identity, so
that the difference vectors that determine the step for any one chain
depends on the history of all three chains. This property allows any
chain in $\mathbf{X}$ to hop between modes in the posterior
distributions of the parameters. To search for modes and allow their
inclusion in the parameter distribution, the algorithm sets $\gamma =
1$ every ten iterations.

The DREAM(ZS) algorithm also employs a more complicated ''snooker''
updater that relies instead on three draws from $\mathbf{Z}$. 
\citet{terbraak08} demonstrated that the snooker updater
complements the difference updater described above, arguably allowing a more
efficient search of parameter space than would be provided by
difference steps alone. In our implementation, there is a 10\% chance
of a given iteration using snooker rather than difference updates.

We run $10^5$ iterations, which, taking into account three chains and
ten steps per interval, corresponds to $3\times 10^6$ functional
evaluations of the model.
Parameter distributions and summary statistics are derived from the
final state of the history array $\mathbf{Z}$. Since we initialize
$\mathbf{Z}$ using random draws from the parameter support regions,
the early values of $\mathbf{Z}$ are drawn far from the equilibrium
parameter distribution and represent expectedly poor fits to the
data. These initial ``burn-in'' values are identified based on an
evaluation of $\chi^2$ values, and they are discarded before analysis
of the chains. Formally, we divide the $\chi^2$ vector into 20 equal
intervals. The boundary between non-equilibrium and equilibrium
intervals is determined using the \citet{geweke92} diagnostic to evaluate
the difference of means between a given interval and the final
interval.

\note{In our earlier {\tt AMOEBA} analysis, we found that the uncertainties in our measured spectrum were systematically smaller than the residuals of the fit.  Adding in an additional $\sim$10\% uncertainty brings the residuals and uncertainties into reasonable agreement.  This discrepancy could be due to  underestimated uncertainties in our observed spectrum and/or systematic uncertainties in our telluric spectrum and the brown dwarf model atmospheres.  To account for the effect of these systematic uncertainties in the posterior distributions of our MCMC analysis, we include a uniform 10\% systematic uncertainty\footnote{Because {\tt AMOEBA} does not propagate uncertainties, including a systematic 10\% uncertainty reduces the $\chi^2$ of our AMOEBA fits, but does not affect the best fit parameters.} which is added in quadrature to the measurement errors of our observed spectrum.}

Our spectrum of \pso\ as well as the best fit model is displayed in Figure \ref{fig:model_fit}.  \note{The best fit model has \teff=1325$^{+330}_{-12}$~K and \logg=3.7$^{+1.1}_{-0.1}$~dex.  We determine a best fit LSF FWHM of 0.00016$\pm$0.00001~$\mu$m, in agreement with the LSF FWHM from our arc lamp spectrum (0.00016$\pm$0.00002~$\mu$m).  The reduced $\chi^2$ for our best fit model is 1.24.  Figure \ref{fig:dreamz_fits} shows the marginalized posterior distributions for $v_r$ and \vsini.  We find $v_r$=$-$6.0$^{+0.8}_{-1.1}$ \kms\ and \vsini = 17.5$^{+2.3}_{-2.8}$~\kms\ for \pso. } 

\section{Membership in $\beta$~Pictoris}

\begin{deluxetable*}{lrrrrrr}
\tablecaption{Galactic Positions and Velocities}
\tabletypesize{\scriptsize}
\tablewidth{0pt}
\tablecolumns{2}
\tablehead{
  \colhead{Name} &
  \colhead{X} &
  \colhead{Y} &
  \colhead{Z} &
  \colhead{U} &
  \colhead{V} &
  \colhead{W} \\
  \colhead{\nodata} &
  \colhead{pc} &
  \colhead{pc} &
  \colhead{pc} &
  \colhead{\kms} &
  \colhead{\kms} &
  \colhead{\kms} 
}
\startdata
\bpic\ Group\tablenotemark{a} & 8.4$\pm$31.9 &-5.0$\pm$15.4 & -15.0$\pm$8.0 & -10.9$\pm$1.5 & -16.0$\pm$1.4 & -9.2$\pm$1.8 \\
\pso & 15.2$\pm$0.6 & 7.2$\pm$0.3 & -14.6$\pm$0.6 & -10.4$\pm$0.7 & -16.4$\pm$0.6 & -9.8$\pm$0.8\\
Offset & 6.7$\pm$31.8 & 12.2$\pm$15.4 & 0.4$\pm$8.0 & 0.5$\pm$1.6 & -0.4$\pm$1.7 & -0.6$\pm$2.0
\enddata

\tablenotetext{a}{From \citet{mamajek14}.  The uncertainties in $UVW$ include both uncertainties in the mean $UVW$ for the group as well as the 1$\sigma$ dispersion of the group members. }

\label{tbl:uvw}
\end{deluxetable*}

Based on its parallactic distance and proper motion, \citet{liu13} identified \pso\ as a likely \bpic\ member.  The Bayesian analysis of \citet{gagne14} found a 99.7\%  probability that \pso\ is a \bpic\ member and predicted a radial velocity of $-6.4\pm$1.7~\kms.  
To determine the full, three-dimensional velocity and position of \pso, we combine our $v_r$ measurement with the updated parallax and proper motion from Liu \etal\ (2015, in preparation): $\pi = 0$\farcs$0450\pm0$\farcs$0017$; $\mu_{\alpha} \rm{cos} \delta = 0$\farcs$1362\pm0$\farcs$0011$~yr$^{-1}$; $\mu_{\delta} = -0$\farcs$1441\pm0$\farcs$0012$~yr$^{-1}$.
Using the three-dimensional velocity and position of \pso, we recalculated membership probabilities using the BANYAN-II web tool\footnote{\url{http://www.astro.umontreal.ca/~gagne/banyanII.php}} \citep{gagne14, malo13} and find that the probability of membership in \bpic\ increases to 99.98\%.  Thus, our measured radial velocity is consistent with membership in \bpic.

\begin{figure*}
\centerline{\includegraphics[angle=90, width=6in]{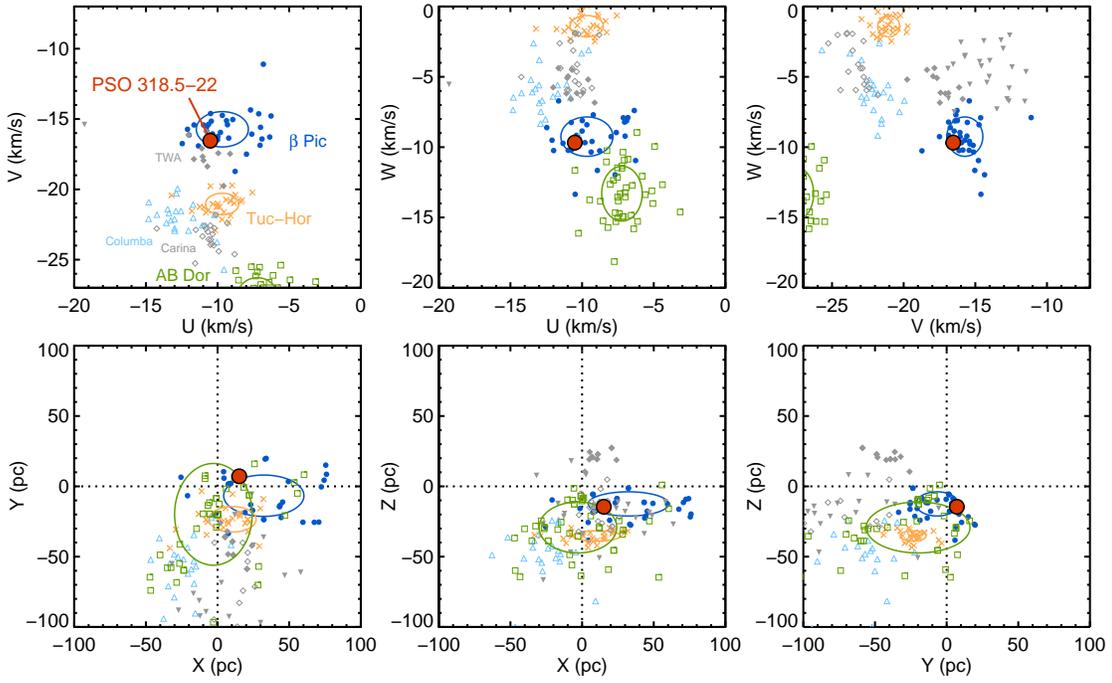}}
\caption[Galactic Position and Velocity]
{\label{fig:uvwxyz} Comparison of the galactic position and velocity of \pso\ to members of known young moving groups from \citet{torres08}.  Ellipses represent the rms for members of the group.  The uncertainties in $UVW$ and $XYZ$ for \pso\ are smaller than the plotted red circle.}
\end{figure*}

We determine galactic $XYZ$ position and $UVW$ velocities for \pso\ using its parallax and proper motions and our radial velocity measurement.  We use a Monte Carlo approach with 10000 iterations to determine uncertainties in our calculated parameters ($UVWXYZ$) based on the uncertainties in our measured quantities.  Table \ref{tbl:uvw} includes the $XYZ$ positions and $UVW$ velocities for both \pso\ and the \bpic\ group \citep{mamajek14}.  As shown in Table \ref{tbl:uvw}, the difference in the galactic position and velocity between \pso\ and the \bpic\ group are well within the uncertainties.  The total velocity of \pso\ is 0.8~$\pm$~1.8~\kms\ from the mean velocity of the \bpic\ group.  The reduced $\chi^2$ for membership based on $UVWXYZ$ and associated uncertainties \note{(Table \ref{tbl:uvw})} for \pso\ and the \bpic\ group is a mere 0.15.  For comparison, we calculated the velocity offset from the mean velocity of the \bpic\ group for each of the stars listed as ``bona fide'' \bpic\ moving group members in \citet{gagne14}\footnote{https://jgagneastro.wordpress.com/banyanii/}. We find that \pso\ is a closer match to the mean kinematics of the \bpic\ group than 46 out of 50 of the \citeauthor{gagne14} "bona fide" members.  Figure \ref{fig:uvwxyz} shows the galactic position and velocities of \pso\ compared to young moving groups from \citet{torres08}.  Regardless of which list of known members is used for comparison \citep{mamajek14, gagne14, torres08}, \pso\ is determined to be a member of the \bpic\ group.

\section{The Physical Properties of \pso}

\begin{deluxetable*}{lcrrrr}
\tablecaption{Physical Properties from Evolutionary Models}
\tabletypesize{\scriptsize}
\tablewidth{0pt}
\tablecolumns{6}
\tablehead{
  \multicolumn{2}{c}{Evolutionary Model} &
  \colhead{Mass} &
  \colhead{\teff} &
  \colhead{Radius} &
  \colhead{\logg} \\
  \colhead{Name} &
  \colhead{Ref} &
  \colhead{\mjup} &
  \colhead{K} &
  \colhead{\rjup} &
  \colhead{dex}
}
\startdata
\sidehead{\underline{\pso\ ($\rm{log}(L_{bol}/L_{\odot}) = -4.52\pm0.04$~dex\tablenotemark{5}; age = 23$\pm$3~Myr\tablenotemark{7})}:}
AMES-COND & 1 &$7.9\pm0.4$ & $1176^{+26}_{-25}$ & $1.358\pm0.010$ & $4.03\pm0.03$\\
AMES-Dusty & 4 &$8.7\pm0.4$ & $1154^{+25}_{-27}$ & $1.417\pm0.007$ &  $4.00\pm0.02$\\
\citeauthor{saumon08} cloudless & 8 & $7.9\pm0.4$ & $1164^{+26}_{-27}$ & $1.373\pm0.010$ &  $4.04\pm0.03$ \\
\citeauthor{saumon08} $f_{sed}$=2 & 8 & $8.3\pm0.5$ & $1127^{+24}_{-26}$ & $1.464\pm0.010$ &  $4.01\pm0.03$ \\
\sidehead{\underline{\bpic~b ($\rm{log}(L_{bol}/L_{\odot}) = -3.86\pm0.04$~dex\tablenotemark{6}; age = 23$\pm$3~Myr\tablenotemark{7})}:}
\citeauthor{saumon08} $f_{sed}$=2 & 8 & $12.8\pm0.2$ & $1583^{+30}_{-33}$ & $1.576\pm0.010$ &  $4.135\pm0.003$ \\
\sidehead{\underline{2M1207b ($\rm{log}(L_{bol}/L_{\odot}) = -4.68\pm0.05$~dex\tablenotemark{2}; age = 10$\pm$3~Myr\tablenotemark{3})}:}
\citeauthor{saumon08} $f_{sed}$=2 & 8 & $4.5\pm0.5$ & $1006^{+24}_{-26}$ & $1.515^{+0.016}_{-0.013}$ &  $3.71\pm0.06$ 
\enddata

\tablerefs{
(1)~\citet{baraffe03};
(2)~\citet{barman11};
(3)~\citet{bell15};
(4)~\citet{chabrier00}; 
(5)~\citet{liu13};
(6)~\citet{males14};
(7)~\citet{mamajek14};
(8)~\citet{saumon08}
}

\label{tbl:mass}
\end{deluxetable*}

\citet{liu13} estimated the mass of \pso\ based on an age of 12$^{+8}_{-4}$~Myr for \bpic.  Recently, revised age estimates indicate that the group is older.  \citet{mamajek14} determined an age of $23 \pm 3$~Myr by combining isochronal ages \citep{mamajek14, malo14} and lithium depletion boundary ages \citep{binks14, malo14}.  For a uniformly-distributed age of $23 \pm 3$~Myr and a normally-distributed luminosity of $\rm{log}(L_{bol}/L_{\odot}) = -4.52\pm0.04$~dex, we determine the mass, \teff, radius and \logg\ of \pso\ using model isochrones (Table \ref{tbl:mass}).  In general, the parameters determined from all four sets of models agree quite well.  The cloud-free evolutionary models (Ames-COND and \citeauthor{saumon08} cloudless) yield slightly lower masses and smaller radii than models with clouds (Ames-DUSTY and \citeauthor{saumon08} $f_{sed} = 2$).  \pso\ has incredibly red $J-K$ colors, indicating a very cloudy atmosphere \citep{liu13}.  \note{The Ames-DUSTY models are not recommended for use with objects having \teff\ cooler than 1700~K \citep{allard01}.}  Thus, we adopt the physical properties as determined by the \citeauthor{saumon08} $f_{sed}=2$ models:  mass=$8.3\pm0.5$~\mjup, \teff=$1127^{+24}_{-26}$~K, radius=$1.464^{+0.009}_{-0.008}$~\rjup, and \logg=$4.01\pm0.03$~dex. 

\note{The \teff\ determined by forward-modeling our spectrum, 1325$^{+330}_{-12}$~K, is significantly higher than the \teff\ inferred by evolutionary models.  This discrepancy is similar to the result found by \citet{liu13} using a much lower resolution spectrum spanning a larger wavelength range.  This hints that atmospheric models over-predict \teff\ when fitting either the broad-band spectral morphology seen in low-resolution spectra or the depths of spectral lines seen in high resolution spectra.}

The mass we determine for \pso\ is somewhat higher than the 6.5$^{+1.3}_{-1.0}$~\mjup\ reported by \citet{liu13}.  This can be attributed to the revised (older) age of the \bpic\ moving group. \note{For comparison to \pso, we calculate the properties of \bpic~b and 2M1207b (Table \ref{tbl:mass}) using the \citeauthor{saumon08} $f_{sed}=2$ evolutionary models and accounting for recent revisions to the ages of the \bpic\ and TW~Hydra moving groups.}  The mass of \pso\ ($8.3\pm0.5$~\mjup) is intermediate to the two known directly-imaged exoplanets around \bpic\ members, \bpic~b and 51~Eri~b \citep[$\sim$2~\mjup;][]{macintosh15}.  \note{In addition, \pso\ is an excellent free-floating analog to \bpic~b and 2M1207b, the only other planetary-mass objects having constraints on their rotation.}

\section{Constraining the Rotation and Inclination of \pso}
Our measurement of \vsini\ = \note{17.5$^{+2.3}_{-2.8}$~\kms\ }allows us to constrain the equatorial velocity and inclination of \pso.  \citet{biller15} report the detection of $J_S$-band variability, likely due to rotational modulation of inhomogeneous cloud cover.  Unfortunately, the measurements of \citet{biller15} do not cover a long enough time span to measure a full rotational period of \pso, but they constrain the period to $\ge$5~hours.  Assuming a radius of \pso\ of 1.464~\rjup, a period of $>$5~hours corresponds to an equatorial velocity of $<$36~\kms.  \note{Thus, our \vsini\ measurement in combination with the \citeauthor{biller15} result constrains the equatorial velocity to 17.5--36~\kms and the inclination of \pso\ to $i \ge 29^{\circ}$.}  In addition, our \vsini\ measurement places an upper limit on the rotational period of \note{10.2~hours}.  A period of 5--10.2 hours is slightly longer than the typical $\sim$3~hour periods of L/T transition field brown dwarfs \citep{radigan14}, which is expected given that \pso\ is young and still in the process of contracting.

\begin{figure}
\centerline{\includegraphics[width=3.5in]{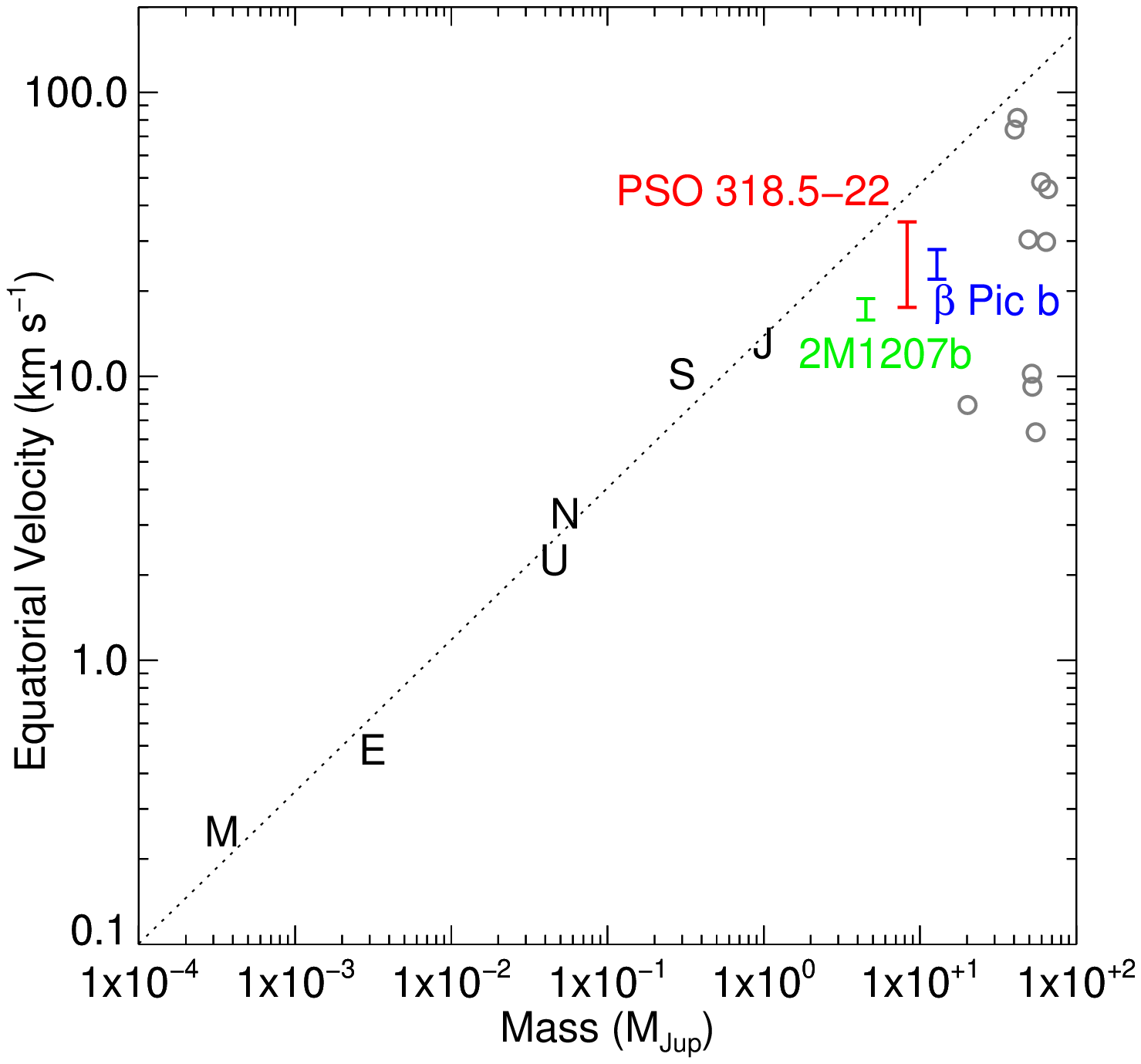}}
\caption[Equatorial Velocities]
{\label{fig:velocity} A variation on Figure 2 of \citet{snellen14} comparing the equatorial velocities of planets and brown dwarfs.  The red bar indicates our constraints on the equatorial velocity of \pso. The blue data point shows the equatorial velocity \citep{snellen14} and estimated mass (see \S 5) of \bpic~b.  \note{The green data point shows the equatorial velocity of 2M1207b calculated from its rotation period \citep{zhou15} and estimated radius (Table \ref{tbl:mass}).}  The gray open circles indicate the equatorial velocities of field brown dwarfs calculated from period measurements in \citet{metchev15} and mass and radius estimates from \citet{filippazzo15}.  Solar system objects are plotted as the first letter of their name.  Mercury and Venus are excluded due to their extremely low equatorial velocities.  The dotted line shows a linear fit to log(mass) vs. log(velocity) for the solar system planets plotted.}
\end{figure}

Our constraints on its equatorial velocity (17.5--36~\kms) allow us to compare the rotation of \pso\ to recent measurements of the rotation of other planetary-mass and substellar objects.  
\note{The equatorial velocity of \bpic~b, 25~\kms \citep{snellen14}, is consistent with our constraint on the equatorial velocity of \pso.  The 10.7$^{+1.2}_{-0.8}$~hour period of 2M1207b \citep{zhou15} is slighty longer than our upper limit on the period for \pso.  Given 2M1207b's lower mass and younger age (see \S 5), it is expected to rotate more slowly than \bpic~b and \pso.  Indeed, the equatorial velocity of 2M1207b is 17.3$\pm$1.5~\kms as calculated from its rotational period and evolutionary-model inferred radius.}  The agreement in equatorial velocities for 2M1207b, \pso\ and \bpic~b hints that the angular momentum evolution of free-floating and gravitationally-bound planetary mass objects is similar.  For comparison, we calculate the equatorial velocities for field brown dwarfs with measured periods in \citet{metchev15} and with estimated masses and radii from \citet{filippazzo15}.  We find that field brown dwarfs have a relatively broad range of equatorial velocities $\sim$6--80~\kms.   \note{Additional rotation measurements for planetary-mass objects are needed to determine if the field brown dwarf spread in equatorial velocity extends to lower masses.}  Figure \ref{fig:velocity} compares the equatorial velocity of \pso\ to that of \bpic~b \citep{snellen14}, planets in our solar system\footnote{\url{http://nssdc.gsfc.nasa.gov/planetary/factsheet/}}, and field brown brown dwarfs.  Solar system planets have a log-log velocity-mass relationship.  2M1207b, \bpic~b and \pso\ are slower rotating than would be expected from the extrapolation of the solar system velocity-mass relationship.  This is likely due to their young ages.  By the time they are the age of our solar system, the equatorial velocities of 2M1207b, \bpic~b, and \pso\ should increase by a factor of $\sim$1.5 as they contract to a radius of $\sim$1~\rjup, putting them in good agreement with the velocity-mass relation seen for the solar system planets.

\section{Conclusions}

Using forward modeling techniques, we have measured the radial velocity ($-$6.0$^{+0.8}_{-1.1}$~\kms) and \vsini\ (17.5$^{+2.3}_{-2.8}$~\kms) of \pso.  Combining our $v_r$ with an updated parallax and proper motion of \pso, we determine its $UVW$ velocities.  We find that the galactic position and motion of \pso\ are consistent with membership in the \bpic\ moving group.  The luminosity of \pso, combined with recent age determinations of the \bpic\ moving group allow us to use evolutionary models to estimate its mass, \teff, radius, and \logg.  Using our \vsini\ measurement and recently published variability results, we determine that the inclination of \pso\ must be greater than $29^{\circ}$ and the rotational period is constrained to 5--10.2~hours.  Our constraints on the equatorial velocity of \pso\ indicate that its rotation is consistent with an extrapolation of the velocity-mass relationship of solar system planets.

With its well-determined age, luminosity, and mass, \pso\ is an important benchmark for studies of young, directly imaged planets.

\acknowledgments
Based on observations obtained via Director's Discretionary Time at the Gemini Observatory, which is operated by the Association of Universities for Research in Astronomy, Inc., under a cooperative agreement 
with the NSF on behalf of the Gemini partnership: the National Science Foundation 
(United States), the National Research Council (Canada), CONICYT (Chile), the Australian 
Research Council (Australia), Minist\'{e}rio da Ci\^{e}ncia, Tecnologia e Inova\c{c}\~{a}o 
(Brazil) and Ministerio de Ciencia, Tecnolog\'{i}a e Innovaci\'{o}n Productiva (Argentina).  This work made use of Bucknell's Linux Computing Cluster maintained by the Engineering Computing Support Team.  KNA and JFG acknowledge support from the Isaac J. Tressler Fund for Astronomy at Bucknell University.

\end{document}